\documentclass[aps,showpacs,twocolumn]{revtex4}
\usepackage{graphicx}
\usepackage{amsmath}
\begin{document}
\title{Simulation of four-body interaction in a nuclear magnetic resonance
quantum information processor \footnote{ Correspondence should be sent to: Gui Lu Long,
gllong @mail. tsinghua. edu.cn}}
\author{\small{} Wen-Zhang Liu$^1$, Jin-Fu
Zhang$^1$, and Gui Lu Long$^{1,2}$}
\address{
$^1$Key Laboratory for Atomic and Molecular NanoSciences and Department of Physics,
Tsinghua University, Beijing 100084, China\\
$^2$  Tsinghua National Laboratory For Information Science and Technology, Beijing
100084, China}
 \pacs{03.67.Lx}
\date{\today}
\begin{abstract}
   Four-body interaction plays an important role in many-body systems, and
   it can exhibit interesting phase transition
behaviors. Historically it was the need to efficiently simulate quantum systems that
lead the idea of a quantum computer. In this Letter, we report the experimental
demonstration of a four-body interaction in a four- qubit nuclear magnetic resonance
quantum information processor. The strongly modulating pulse is used to implement spin
selective excitation. The results show a good agreement between theory and experiment.
\end{abstract}
\pacs{03.67.Lx}
\maketitle

\section{Introduction}
Quantum computers have advantages over the classical counterparts in simulating quantum
systems \cite{Feynman} and solving some hard problems, such as factoring large number
and searching unsorted databases \cite{shor}. Among the various candidates for
implementing the large- scale quantum computer in the future and demonstrating quantum
algorithms to corroborate existing theories, liquid nuclear magnetic resonance (NMR)
has been proven a convenient and practical method to learn lessons for the other
physical systems \cite{book}. NMR quantum computer is an Ising-type computer
\cite{Bowdrey} where two-body interactions take the form of $\pi
J_{ij}\sigma_{z}^{i}\sigma_{z}^{j}/2$, known as $J-$ couplings in NMR, where
$\sigma_{z}^{i}$ denotes the Pauli matrix of the $i$-th spin, and $J_{ij}$ denotes the
strength of coupling between two spins.

There have been much interests in many-body interactions. Besides two-body
interactions, many-body interactions are valuable sources for quantum information
processors. For example, the three-spin interactions can speed up the quantum state
transfer in the Heisenberg spin chain \cite{zhang06}.  Four-body interactions have
attracted much interests recently\cite{four}. The systems with many-body interactions
can exhibit interesting phase transition behaviors \cite{four}, such as quantum
entanglement phase transitions \cite{Yang}.  The three spin-interactions in the spin
chain can induce the quantum criticality that cannot be measured by concurrence,
because three-spin interactions generate three-qubit entanglement \cite{Yang}.
Four-body interaction may play an important role in phase transition in some condensed
matters\cite{four}.

Simulation of quantum system is one of main applications of future quantum computers.
Practical factoring and searching applications of quantum computer usually require
hundreds even thousands of qubits. But the simulation of quantum systems may require
only a few dozens of qubits. Thus simulating quantum systems may well be the first
practical application of the early practical quantum computer.  It is helpful now to
study the simulations of quantum system with a few qubit quantum information processer,
to locate problems and gather experiences, in particular the unitary operations and the
extent of decoherence in existing apparatus. In fact, it has been successfully
demonstrated that three-body interaction can be simulated very well in NMR quantum
computers\cite{zhang06,Tseng}.
  It is a reasonable assumption that the four-spin
interactions relate to four-qubit entanglement, which is still unclear for us
currently. In this paper we focus on implementing the four-spin interactions in a NMR
quantum computer. Our work is a valuable step in exploring quantum simulations, and
also a crucial step for implementing the quantum phase transitions induced by four-body
interactions and investigating the relation between the four-spin interactions and
four-qubit entanglement experimentally.

\section{Generating four-spin interactions using NMR}
  Our task is to decompose the four- spin evolution into
a series of one-spin operations and $J$-couplings. The one-spin
operations are realized by radio frequency pulses. The $J$-coupling
evolution
\begin{equation}\label{zz}
  U_{zz}(T)=e^{-i
  (\pi/2)J_{ij}T\sigma_{z}^{i}\sigma_{z}^{j}}
\end{equation}
can be realized by standard NMR spin-echo techniques \cite{Linden}.
Under $U_{zz}(T)$, $\sigma_x^i$ evolutes as
\begin{equation}
 \sigma_x^i\xrightarrow{J_{ij}}\sigma_x^i\cos \theta +\sigma_y^i\sigma_z^j\sin \theta£¬
\end{equation}
where $\theta=\pi J_{ij} t$.

Through some calculations \cite{decompose} one finds that the $n$-
spin interaction can be decomposed as the $(n-1)$- spin interactions
by iteration
\begin{eqnarray}\label{iteral}
&&e^{-i (\pi/2)J_{12\cdots n}T
\sigma_{z}^{1}\sigma_{z}^{2}\cdots\sigma_{z}^{n}} \nonumber\\
&=&e^{-i (\pi/4)\sigma_{x}^2}e^{-i (\pi/4)\sigma_{z}^1\sigma_{z}^2}\nonumber\\
&&\times e^{-i (\pi/4)\sigma_{y}^2}e^{-i (\pi/2)J_{12\cdots n}T
\sigma_{z}^{2}\cdots\sigma_{z}^{n}}\nonumber\\
&&\times e^{-i (\pi/4)\sigma_{y}^2}e^{-i (\pi/4)\sigma_{z}^1\sigma_{z}^2}e^{i
(\pi/2)\sigma_{y}^2}e^{i (\pi/4)\sigma_{x}^2}.
\end{eqnarray}
By introducing
\begin{eqnarray}
P_{1}(n)&=&e^{-i (\pi/4)\sigma_{x}^{n+1}}e^{-i
(\pi/4)\sigma_{z}^n\sigma_{z}^{n+1}}e^{-i (\pi/4)\sigma_{y}^{n+1}},\label{p1}\\
P_{2}(n)&=&e^{-i (\pi/4)\sigma_{y}^{n+1}}e^{-i (\pi/4)\sigma_{z}^n\sigma_{z}^{n+1}}e^{i
(\pi/2)\sigma_{y}^{n+1}}e^{i (\pi/4)\sigma_{x}^{n+1}}.\nonumber\\ \label{p2}
\end{eqnarray}
Eq. (\ref{iteral}) can be further expressed as
\begin{eqnarray}
&& e^{-i (\pi/2)J_{12\cdots n}T
\sigma_{z}^{1}\sigma_{z}^{2}\cdots\sigma_{z}^{n}}=\nonumber\\
&&\prod_{l=1}^{n-2}P_{1}(l)e^{-i (\pi/2)J_{12\cdots n}T
\sigma_{z}^{n-1}\sigma_{z}^{n}}\prod_{m=1}^{n-2}P_{2}(n-1-m).\nonumber\\
\label{fiteral}
\end{eqnarray}
From Eq. (\ref{fiteral}) one finds that the many-  spin interaction can
be decomposed into the operations that can be directly realized by NMR. When $n=4$ one
obtains
\begin{equation}\label{Uzzzz2}
\begin{split}
U_{zzzz}(T)&= e^{-i (\pi/2)J_{1234}T
\sigma_{z}^{1}\sigma_{z}^{2}\sigma_{z}^{3}\sigma_{z}^{4}}\\
 &=e^{-i
(\pi/4)\sigma_{x}^2}e^{-i (\pi/4)\sigma_{z}^1\sigma_{z}^2}e^{-i
(\pi/4)\sigma_{y}^2}e^{-i (\pi/4)\sigma_{x}^3}\\
&\times e^{-i (\pi/4)\sigma_{z}^2\sigma_{z}^3}e^{-i (\pi/4)\sigma_{y}^3}e^{-i(\pi/2)
J_{1234} T \sigma_z^3\sigma_z^4}\\
&\times e^{-i (\pi/4)\sigma_{y}^3}e^{-i (\pi/4)\sigma_{z}^2\sigma_{z}^3}e^{i
(\pi/2)\sigma_{y}^3}e^{i (\pi/4) \sigma_{x}^3}\\
&\times e^{-i (\pi/4)\sigma_{y}^2}e^{-i (\pi/4)\sigma_{z}^1\sigma_{z}^2}e^{i
(\pi/2)\sigma_{y}^2}e^{i (\pi/4)\sigma_{x}^2},\\
\end{split}
\end{equation}
where $J_{1234}$ is the effective strength of the four- spin interaction. The above
equation can also be represented as
\begin{equation}\label{Uzzzz1}
\begin{split}
U_{zzzz}(T)&=e^{-i (\pi/4)\sigma_{x}^2}e^{-i (\pi/4)\sigma_{z}^1\sigma_{z}^2}e^{-i
(\pi/4)\sigma_{y}^2}e^{-i (\pi/4)\sigma_{x}^2}\\
&\times e^{-i (\pi/4)\sigma_{z}^2\sigma_{z}^3}e^{-i (\pi/4)\sigma_{y}^2}e^{-i(\pi/2)
J_{1234} T \sigma_z^2\sigma_z^4}e^{i (\pi/4)\sigma_{y}^2}\\
&\times e^{-i (\pi/2)\sigma_{y}^2}e^{-i (\pi/4)\sigma_{z}^2\sigma_{z}^3}e^{i
(\pi/2)\sigma_{y}^2}e^{i (\pi/4)\sigma_{x}^2}\\
&\times e^{-i (\pi/4)\sigma_{y}^2}e^{-i (\pi/4)\sigma_{z}^1\sigma_{z}^2}e^{i
(\pi/2)\sigma_{y}^2}e^{i (\pi/4)\sigma_{x}^2}.\\
\end{split}
\end{equation}

\section{implementation}
\begin{figure}
\includegraphics[width=2in]{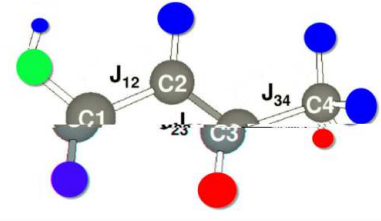}
\caption{(Color online) Molecule of crotonic acid. The blue spheres denote protons and
the green spheres denote oxygens. The chemical shifts are $\nu_1=$21468.9 Hz, $\nu_2=$
15255.6 Hz, $\nu_3=$ 18668.0 Hz, and $\nu_4=$ 2190.4 Hz. The $J$ coupling constants are
$J_{12}=$72.4 Hz, $J_{13}=$-1.3 Hz,$ J_{14}=$7.0 Hz, J$_{23}=$70.3 Hz, $J_{24}=$-1.6
Hz, and $J_{34}=$41.3 Hz. } \label{fig1}
\end{figure}

  We use Carbon-13 labelled crotonic acid dissolved in D$_{2}$O as the sample.
  The chemical sketch of crotonic acid is shown as Fig. \ref{fig1}, where C1 - C4 are
assigned as qubits 1 - 4, respectively. The protons are decoupled during the whole
experiment. The experiments are implemented on a Bruker DRX 500 MHz spectrometer. The
temperature is controlled at 22 $^{\circ}$C.

\begin{center}
\begin{figure}[h]
\includegraphics[angle=270, width=3.5in]{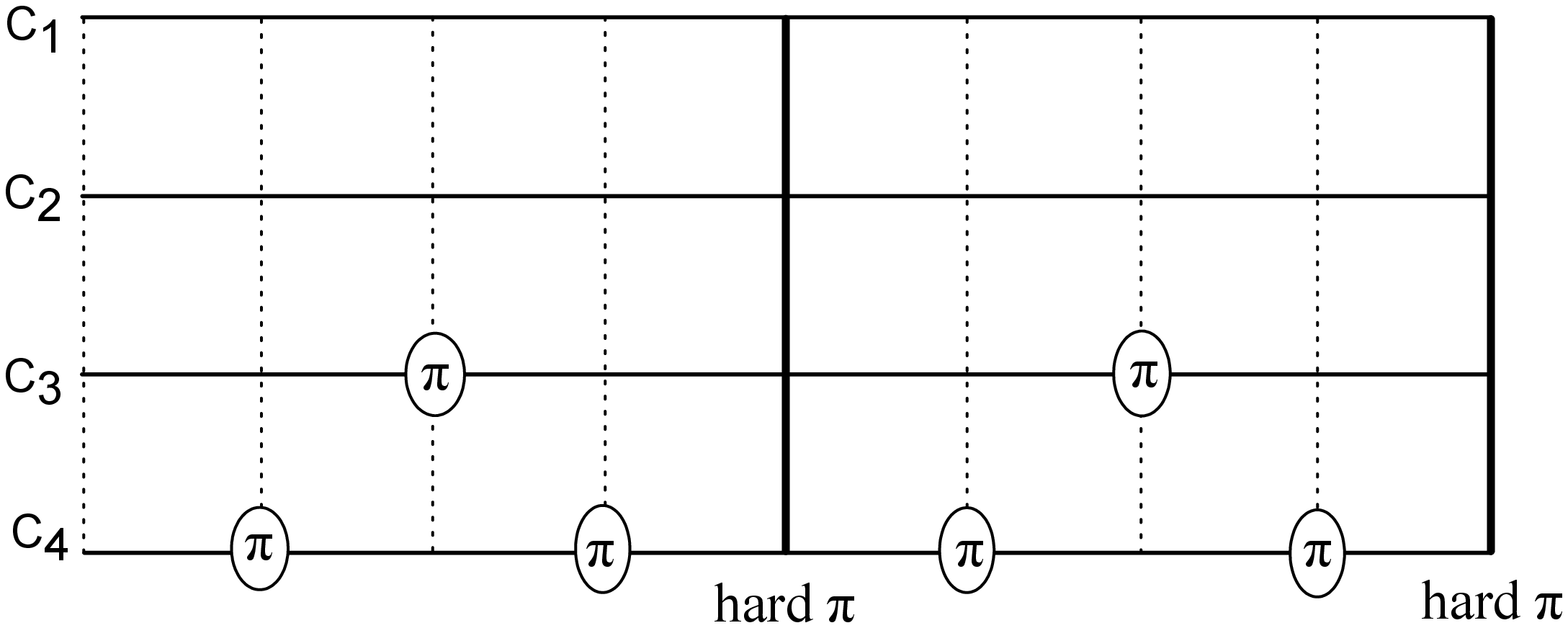}
\caption{Refocusing scheme to implement $[1/2J_{12}]$. The ellipses represent the
strongly modulating $\pi$ pulses. The evolution time $1/2J_{12}$ is divided into eight
identical segments.}\label{fig2}
\end{figure}
\end{center}

The Hamiltonian of the NMR system reads
\begin{equation}\label{nmr}
  H_{NMR}=-\pi\sum_{k=1}^{4}\nu_{k}\sigma_{z}^{k}
  +\frac{1}{2}\pi \sum_{k<l}J_{kl}\sigma_{z}^{k}\sigma_{z}^{l},
\end{equation}
where $\nu_{1}$ -  $\nu_{4}$ are the resonance frequencies of C1 -
 C4. The
coupled-spin evolution between two spins is denoted as
\begin{equation}\label{2}
  [\tau_{kl}]=e^{-i\frac{1}{2}\pi J_{kl} \tau \sigma_{z}^{k} \sigma_{z}^{l}},
\end{equation}
where $k/l=1,2,3,4$, and $k\neq l$. $[\tau_{kl}]$ can be realized by
averaging the coupling constants other than $J_{kl}$ to zero
\cite{Linden}. The pulse sequence to implement $[1/2J_{12}]$ is
shown in Fig.\ref{fig2}, 
 where the evolution time $1/2J_{12}$ is
divided into eight identical segments.

We use Eq. (\ref{Uzzzz2}) to implement $U_{zzzz}(T)$ through the
pulse sequence
\begin{widetext}
\begin{equation}\label{se}
\begin{split}
&[\frac{\pi}{2}]_{x}^{2}\rightarrow[\pi]_{y}^{2}\rightarrow[\frac{1}{2J_{12}}]\rightarrow[\frac{\pi}{2}]_{-y}^{2}
\rightarrow[\frac{\pi}{2}]_{x}^{3}
\rightarrow[\pi]_{y}^{3}\rightarrow[\frac{1}{2J_{23}}]\rightarrow[\frac{\pi}{2}]_{-y}^{3}\rightarrow[\frac{J_{1234}T}{J_{34}}]\\
&\rightarrow[\frac{\pi}{2}]_{-y}^{3}\rightarrow[\frac{1}{2J_{23}}]
\rightarrow[\frac{\pi}{2}]_{-x}^{3}\rightarrow[\frac{\pi}{2}]_{-y}^{2}\rightarrow[\frac{1}{2J_{12}}]\rightarrow[\frac{\pi}{2}]_{-x}^{2}
\end{split}
\end{equation}
\end{widetext}
where $[\frac{\pi}{2}]_{-y}^{2}$ denotes a $\pi/2$ pulse along $-y$ axis on C2. The
corresponding evolution is $e^{-i(\pi/4)\sigma_{y}^{2}}$. All spin selective pulses are
strongly modulated pulses (SMPs) \cite{cory2002}. A SMP consists of a series of
non-selective (hard) pulses that modulate the system's dynamics strongly to produce
precisely a desired spin-selective unitary propagator. In our experiments the fidelity
of each SMP is larger than 0.99. The total duration time of the whole experiment is
about $80$ ms.

\begin{center}
\begin{figure}[h]
\includegraphics[angle=-90,width=3.5in]{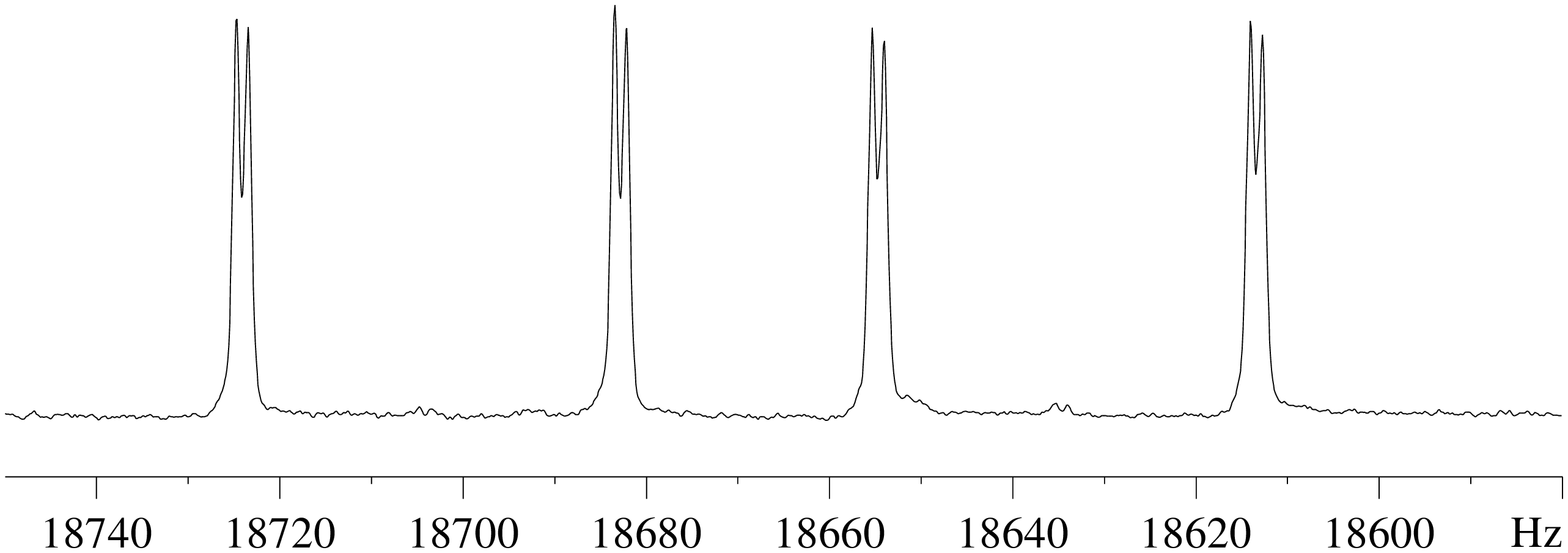}
\caption{The C3 spectra when the system lies in $\rho_{ini}=\sigma_{x}^{3}$.}
\label{fig3}
\end{figure}
\end{center}

 We choose the state
\begin{equation}
\rho_{ini}=\sigma_{x}^{3},
\end{equation}
as the initial state, which is prepared by $$[\frac{\pi}{2}]_{y}^{1,2,4}
\rightarrow[grad]_{z}\rightarrow [\frac{\pi}{2}]_{y}^{3}$$ from the thermal equilibrium
\cite{Tseng,zhang06,zhang05}. Here we use the deviation density matrix to describe the
state of the NMR system \cite{Chuang,fplong}. The carbon spectrum for the system in
$\rho_{ini}$ is shown in Fig. \ref{fig3}, where the signals are chosen as the reference
signals for the following spectra. The small $J_{13}$ causes the partial overlapping
peaks.

\begin{center}
\begin{figure}
\includegraphics[width=3.5in]{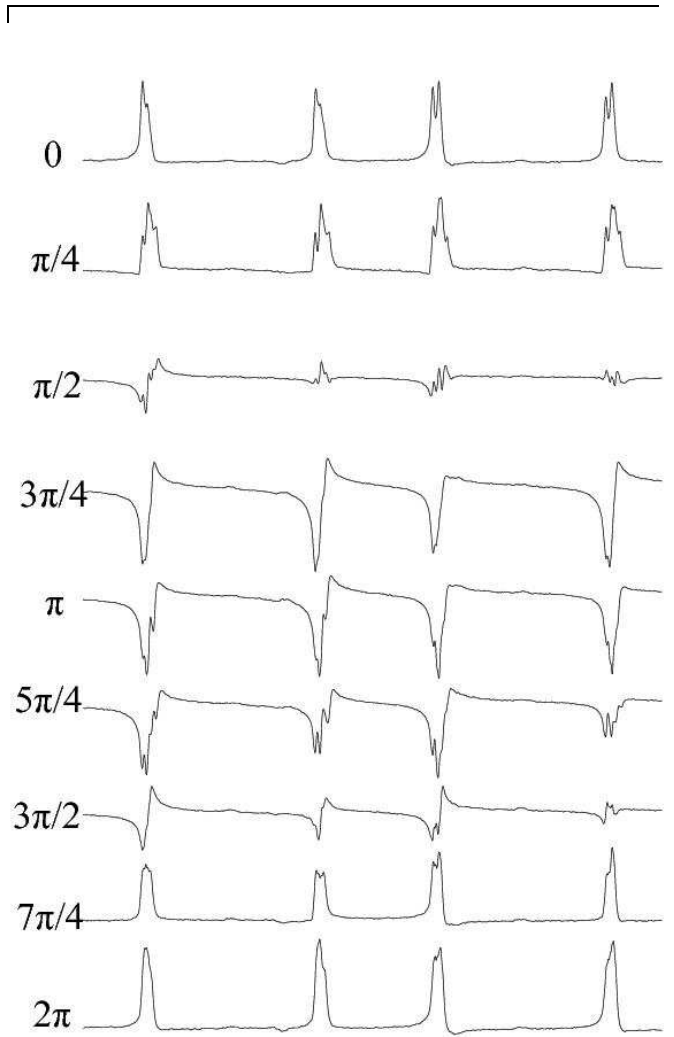}
\caption{NMR spectra of C3 in evolution under a four-body interaction. From top to
bottom, the figures correspond to $\pi J_{1234}T=n \pi/4$ with $n=0$, $1$, ..., $8$,
respectively.}\label{fig4}
\end{figure}
\end{center}

\begin{center}
\begin{figure}
\includegraphics[angle=270,width=3.5in]{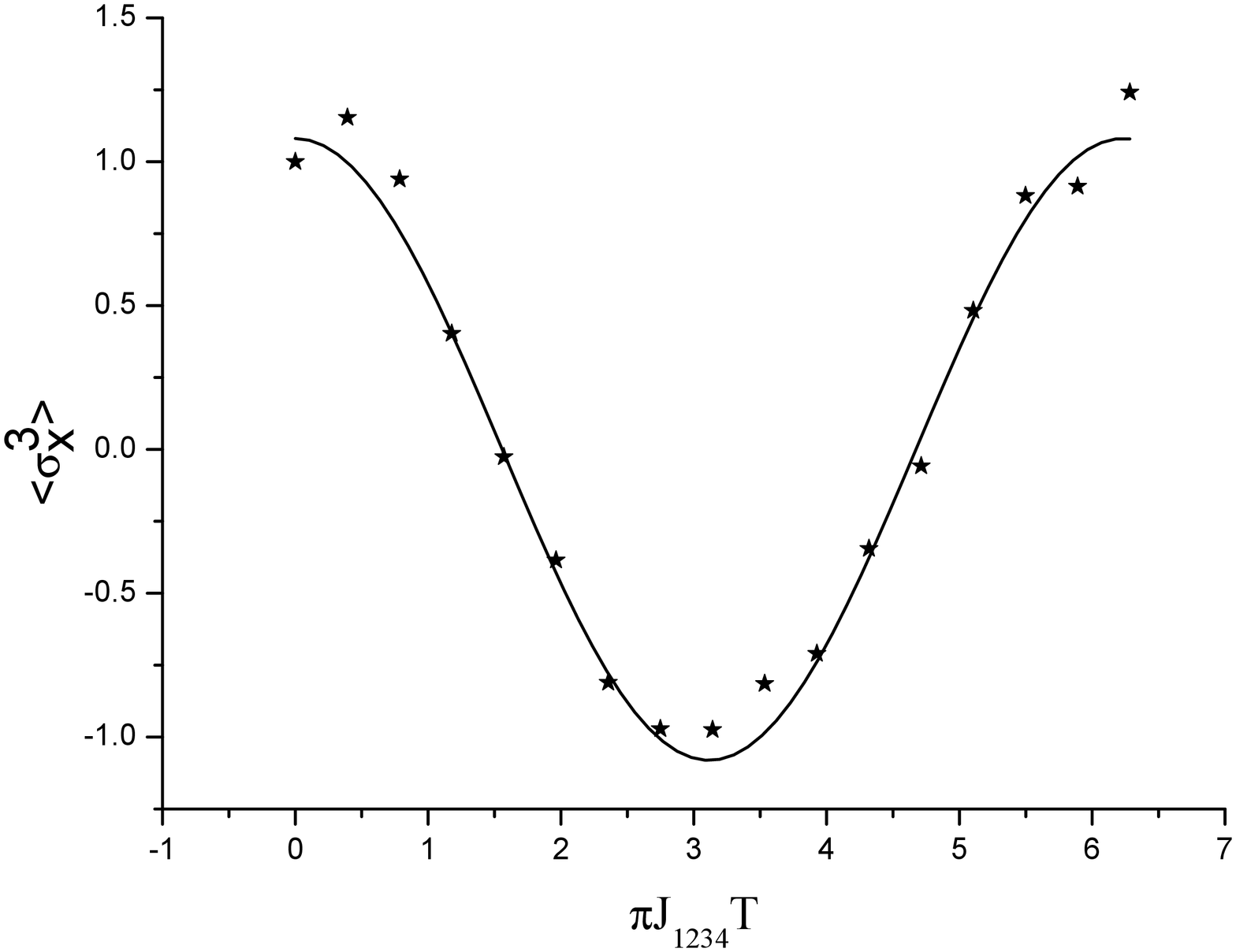}
\caption{The expectation value $\langle \sigma_{x}^{3}\rangle$ versus $\pi J_{1234}T$
under U$_{zzzz}$. The curve can be fitted as $\langle \sigma_{x}^{3}\rangle=1.081
\cos(1.008\pi J_{1234}T)$.} \label{fig5}
\end{figure}
\end{center}

 Under the four-body interaction, the state of this system changes from
$\rho_{ini}$ to
\begin{equation}\label{reszzzz}
\rho_{zzzz}(T)= \sigma_{x}^{3}\cos(\pi
J_{1234}T)+\sigma_{z}^{1}\sigma_{z}^{2}\sigma_{y}^{3}\sigma_{z}^{4}\sin(\pi
J_{1234}T).
\end{equation}
When $\pi J_{1234}T=n \pi/4$ with $n=0$, $1$, ..., $8$, the spectra of C3 are shown as
Fig. \ref{fig4}. The experimental results agree on the theoretical expectations. By
integrating over the eight peaks we obtain the evolution of $\langle
\sigma_{x}^{3}\rangle$ as a function of time, $\pi J_{1234}T$, and is  shown in Fig.
\ref{fig5}. The curve can be fitted as $\langle \sigma_{x}^{3}\rangle=1.081
\cos(1.008\pi J_{1234}T)$, which agrees well with the theoretical expectation $\langle
\sigma_{x}^{3}\rangle= \cos(\pi J_{1234}T)$. The small discrepancy is due to the
imperfection of pulse and decoherence.


\section{Summary}
We have experimentally simulated the four-body interaction in a four-qubit NMR quantum
information processor. The experiment results show good agreement with the theoretical
expectations. The SMP makes the simulation in NMR very well. With this experiment, one
can proceed to demonstrate large quantum system simulations, and look into interesting
physical phenomenon such as phase transitions in quantum systems with four-body
interaction.  The simulation method used here in NMR techniques can be generalized to
other Ising type quantum computer.

\section{Acknowledgment}
 The experiments were performed at physic department of Dortmund University.
We thank the support given by Prof. D. Suter. Liu thanks Dr. T. S. Mahesh for his help
in SMPs. This work is supported by the National Natural Science Foundation of China
under Grant No. 10374010, 60433050, 10325521, the National Fundamental Research Program
Grant No. 2006CB921106, the Hang-Tian Science Fund, the SRFDP program of Education
Ministry of China.




\begin{thebibliography}{}
\bibitem{Feynman}R. P. Feynman, Int. J. Theor. Phys. {\bf 21}, 467
        (1982); S. Lloyd, Science {\bf 273}, 1073 (1996).

\bibitem{shor} P. W. Shor, in Proceedings of the 35th Annual Symposium on
     the Foundations of Computer Science, Santa Fe, NM, 1994 (IEEE
     Computer Society Press, New York 1994); L. K. Grover, Phys.
        Rev. Lett. {\bf 79}, 325 (1997)

\bibitem{book} M. A. Nielsen and I. L. Chuang, {\it Quantum Computation and
                    Quantum Information}, Page 353 (Cambridge University Press, Cambridge,
                    2000); {\it The Physics of Quantum Information}, edited by
                D. Bouwmeester, A. Ekert, and A. Zeilinger, Page 221 (Springer, Berlin,
                2000); D. G. Cory, A. Fahmy and T. F.
                     Havel, Proc. Natl. Acad. Sci. USA {\bf 94},1634(1997);
                N. Gershenfeld and I. Chuang, Science {\bf 275},350(1997);
                    D. G. Cory, M. D. Price, and T. F. Havel, Physica D {\bf 120}, 82
                (1998);
                 L. M. K. Vandersypen and I. L. Chuang, Rev. Mod. Phys. {\bf 76}, 1037
                       (2004)




 \bibitem{Bowdrey}  M. D. Bowdrey, J. A. Jones, E. Knill, R.
                    Laflamme, Phys. Rev. A {\bf 72}, 032315 (2005)

\bibitem{zhang06} J.-F. Zhang, X.-H. Peng, D. Suter, Phys. Rev. A {\bf 73}, 062325
                    (2006)

\bibitem{four} O. G. Mouritsen, B. Frank, D. Mukamel, Phys. Rev. B {\bf 27},
                        3018 (1983);
                V. Mastropietro, Journal of Statistical Physics, {\bf111}, 201 (2003);
                S. S. Aplesnin and N. I. Piskunova, J. Phys.: Condens. Matter {\bf 17}, 5881
                    (2005);
                    A. Lipowski, Physica A {\bf 248}, 207 (1997);
                    T. Iwashita, K. Uragami, K. Goto, T. Kasama, T. Idogaki, Physica B {\bf 329}¨C{\bf 333}, 1284
                    (2003);
                    B. Boechat, et al., Phys. Rev. B {\bf 61}, 14327
                    (2000);
                    D. F. Styer, M. K. Phani, and J. L. Lebowitz, Phys. Rev. B {\bf 34}, 3361
                    (1986).


\bibitem{Yang} P. Lou, W.-C. Wu, and M.-C. Chang, Phys. Rev. B {\bf 70}, 064405 (2004);
                M.-F. Yang, Phys. Rev. A, {\bf 71}, 030302(R)
                    (2005)

\bibitem{Tseng} C. H. Tseng, et al ,Phys. Rev.A 61, 012302 (1999)

\bibitem{Linden} R. R. Ernst,G.Bodenhausen, and A.Wokaum, Principles of
            Nuclear Magnetic Resonance in One and Two Dimensions (Oxford
            University Press,Oxford,1987);
             N. Linden, et al, Chem. Phys. Lett.
                {\bf 311}, 321 (1999);
                J.-F. Zhang, G. L. Long, Z.-W. Deng, W.-Z. Liu, and Z.-H
                Lu, Phys. Rev. A {\bf 70}, 062322 (2004)

\bibitem{decompose} R. Somma, G. Ortiz, J. E. Gubernatis, E. Knill, and R.
        Laflamme, Phys. Rev. A {\bf 65}, 042323 (2002);
        M. D. Price, S. S. Somaroo, A. E. Dunlop, T. F. Havel, and D. G.
        Cory, Phys. Rev. A {\bf 60}, 2777 (1999);
        Shyamal S. Somarooa, David G. Coryb and Timothy F. Havel, Phys. Lett.
            A {\bf 240}, 1 (1998)


\bibitem{cory2002}E. M. Fortunato,M. A. Pravia,N. Boulant,G.
Teklemariam,T. F. Havel and D. G. Cory, Joural of Chemical
Physics,{\bf 116} 7599 (2002); N. Boulant, K. Edmonds, J. Yang, M.
A. Pravia, and D. G. Cory, Phys. Rev. A {\bf 68}, 032305 (2003); C.
Negrevergne, T. S. Mahesh, C. A. Ryan, M. Ditty, F. Cyr-Racine, W.
Power, N. Boulant, T. Havel, D. G. Cory, and R. Laflamme, Phys. Rev.
Lett. {\bf 96}, 170501 (2006)


\bibitem{zhang05} J.-F. Zhang, G. L. Long, W. Zhang, Z.-W. Deng, W.-Z. Liu, and
                  Z.-H. Lu, Phys. Rev. A {\bf 72}, 012331(2005)

\bibitem{Chuang} I. L. Chuang, N. Gershenfeld, M. G. Kubinec, and
D.W. Leung,Proc. R. Soc. London, Ser.A 454,447(1998)

\bibitem{fplong} G. L. Long, Y. F. Zhou, J. Q. Jin, Y. Sun and H. W. Lee, Foundations
of Physics, 36 (8) 1217-1246 (2006)




\end{thebibliography}
\end{document}